\documentstyle[amssymb,aps,twocolumn]{revtex}

\begin{document}
\draft
\title{Peltier effect in normal metal-insulator-heavy fermion metal junctions.}
\author{A. V. Goltsev}
\address{Ioffe Physico-Technical Institute, 194021 St. Petersburg, Russia}
\author{D. M. Rowe, V. L. Kuznetsov, L. A. Kuznetsova, and Gao Min}
\address{NEDO Laboratory for Thermoelectric Engineering, Cardiff University, UK}
\maketitle

\begin{abstract}
A theoretical study has been undertaken of the Peltier effect in normal
metal - insulator - heavy fermion metal junctions. The results indicate
that, at temperatures below the Kondo temperature, such junctions can be
used as electronic microrefrigerators to cool the normal metal electrode and
are several times more efficient in cooling than the normal metal - heavy
fermion metal junctions.
\end{abstract}




The attention of physicists has been drawn to heavy fermion (HF) compounds 
\cite{gs91,h93} the unusual properties of which make them attractive for
various physical applications such as thermoelectric devices \cite
{m97,mb99,rs00}. In this paper the Peltier effect of a normal metal (N) -
insulator (I) - HF metal (N$_{HF}$) junction is investigated. It is shown
that NIN$_{HF}$ junction can be used as an electronic microrefrigerator. The
physical principle of the microrefrigerator is that the current flowing
through the junction removes high-energy thermal electrons from the normal
metal, thus cooling it. Recently, electronic microrefrigerators based on a
normal metal-insulator-superconductor (NIS) tunnel junctions have been
proposed and investigated \cite{nem94,ba95}. It is the existence of a
superconducting gap in the superconductive electrode of the NIS junction
that permits manipulation of the Fermi-Dirac distribution of electrons.
Similar refrigeration effect also can be achieved through the internal field
emission in a thin-film device \cite{mc99}, vacuum devices \cite{hgmk01,f01}
and superconductor-semiconductor-superconductor structures \cite{spk01}.

In the present paper it is shown that the existence of a hybridization gap
produced by strong electron correlations near the Fermi surface in a HF
metal also allows manipulation of the energy transfer from the electron gas
in the normal metal electrode (thermoelement) to electrons in the HF
electrode in the NIN$_{HF}$ junction. Calculation of the temperature
dependence of the power transfer from the normal metal electrode reveals
that it attains a maximum at a temperature below the Kondo temperature $%
T_{k} $.

The tunneling current and thermal properties of the NIN$_{HF}$ junction
shown in Fig. \ref{figure1}(a) is investigated. A potential $V$ is applied to
the normal metal electrode. The tunneling current is given by

\begin{eqnarray}
I &=&2\pi e\sum_{\nu \alpha k,\sigma p}\left| T_{\nu \alpha k,\sigma
p}\right| ^{2}\delta (\varepsilon _{Lp}+eV-E_{\nu k})\times  \nonumber \\
&&[f(\varepsilon _{Lp})-f(E_{\nu k})]  \label{current1}
\end{eqnarray}
where $\varepsilon _{Lp}$ is the kinetic energy of conduction electrons with
momentum $p$ in the normal metal electrode, $f(E)$ is the Fermi-Dirac
distribution function,$\ T_{\nu \alpha k,\sigma p}=\left\langle \nu \alpha
k\mid H\mid \sigma p\right\rangle $ is the matrix element of the electron
transition through the barrier from an electron state ($\sigma p)$ in the
normal metal electrode into a quasiparticle state ($\nu \alpha k$) in the
energy band $E_{\nu }({\bf k})$ of the HF metal, $\sigma $ and $\alpha $ are
spin indices, $H$ is the Hamiltonian of the junction.

The thermal transport properties of the NIN$_{HF}$ junction are determined
by the energy transferred by the tunneling current. When electrons with
kinetic energy $\varepsilon _{Lp}$ are removed from the normal metal and
transit into the HF metal, then electrons with an average kinetic energy
equal to the chemical potential $\mu $ are returned to the normal metal
electrode through normal contact. Then the power transfer $P$\ from the
normal metal electrode on the left into the HF electrode on the right is
given by

\begin{eqnarray}
P &=&2\pi \sum_{\nu \alpha k,\sigma p}\left| T_{\nu \alpha k,\sigma
p}\right| ^{2}\delta (\varepsilon _{Lp}+eV-E_{\nu k})\times  \nonumber \\
&&(\varepsilon _{Lp}-\mu )[f(\varepsilon _{Lp})-f(E_{\nu k})].
\label{power1}
\end{eqnarray}

If $P>0$ then the tunneling current cools the electron gas in the normal
metal electrode. In order to determine $I$ and $P$ it is necessary to obtain 
$T_{\nu \alpha k,\sigma p}$ and $E_{\nu }({\bf k}).$ This is a non-trivial
task for HF compounds due to the strong electron correlations. The physical
properties of HF compounds depend strongly on temperature due to the Kondo
effect. At temperatures $T>>T_{k}$ bare electron states are well defined
quasiparticle states. With decreasing temperature when $T\rightarrow T_{k}$,
the scattering of conduction electrons off the localized $f$ electrons is
enhanced and results in the Kondo resonance which is responsible for the
unusual transport properties of HF compounds \cite{bcw87}. In particular, at 
$T\sim T_{k}$ the Seebeck coefficient of a HF metal displays a giant peak.\
At temperatures below $T_{k}$ hybrid quasiparticles are formed. They can be
described as a quantum superposition of conduction and $f$ electrons. Thus,
it is expected that tunneling of electrons from the normal metal electrode
into the HF electrode is different above and below $T_{k}$ due to the
different nature of quasiparticle states.

In order to study the Peltier effect of the NIN$_{HF}$ junction at $T<T_{k}$
the following Hamiltonian was considered: $H=H_{L}+H_{R}+H_{T}$ where $H_{L}$
and $H_{R}$ are the Hamiltonians of the normal and HF metals, respectively. $%
H_{T}$ is the tunneling Hamiltonian:

\begin{equation}
H_{T}=\sum_{k,p,\sigma }(T_{k,p}a_{\sigma p}^{+}c_{\sigma k}+T_{k,p}^{\ast
}c_{\sigma k}^{+}a_{\sigma p})  \label{tunneling}
\end{equation}
where $a_{\sigma p}^{+}$$(a_{\sigma p})$ and $c_{\sigma k}^{+}$($c_{\sigma
k})$ are creation (annihilation) operators for bare electron states with
spin $\sigma $ and momentum $p$ and $k$ in the normal and HF metals,
respectively, $T_{k,p}$ is the bare matrix element. $H_{R}$ is the
Hamiltonian of the periodic Anderson model. At $T<T_{k}$, in the framework
of the mean field approach, the Hamiltonian $H_{R}$ can be written in the
form \cite{nr87,Ikeda96,mc00}:

\begin{eqnarray}
H_{R} &=&\sum_{\sigma k}\varepsilon _{Rk}c_{\sigma k}^{+}c_{\sigma
k}+\sum_{\alpha k}\varepsilon _{f}f_{ak}^{+}f_{ak}  \nonumber \\
&&+\sum_{\sigma \alpha k}(V_{eff}\phi _{\alpha \sigma }({\bf k)}%
f_{ak}^{+}c_{\sigma k}+H.c.)  \label{Anderson}
\end{eqnarray}
where $\varepsilon _{Rk}$ is the kinetic energy of noninteracting conduction
electrons in the HF metal, $V_{eff}$\ and $\varepsilon _{f}$ are the
effective hybridization parameter and effective $f$-level energy,
respectively. The parameters $V_{eff}$, $\varepsilon _{f}$ and the chemical
potential $\mu $ have to be determined self-consistently, minimizing the
free energy with respect to $V_{eff}$, $\varepsilon _{f}$ and $\mu $ at a
given total number $N_{t}$ of electrons. Here $N_{t}=N_{c}+N_{f}$ where $%
N_{c}$ and $N_{f}$ are the number of conduction and $f$ electrons per $f$
ion, respectively. This gives a set of the mean-field equations (see, for
example, \cite{nr87}). Ce$^{+3}$ ions in the ground state have one electron
in the $f$-shell. Valence fluctuations leads to $N_{f}\lesssim 1$. In this
case $\varepsilon _{f}>\mu $. \ The function $\phi _{\alpha \sigma }({\bf k)}
$ characterizes the anisotropy of the Kondo coupling. The diagonalization of
the Hamiltonian (\ref{Anderson}) gives the quasiparticle energies

\begin{equation}
E_{\nu k}=\{\varepsilon _{f}+\varepsilon _{Rk}\mp \lbrack (\varepsilon
_{f}-\varepsilon _{Rk})^{2}+4\left| V({\bf k})\right| ^{2}]^{1/2}\}/2.
\label{h-bands}
\end{equation}
Here $V({\bf k})\equiv V_{eff}\phi ({\bf k)}$, $\phi ^{2}({\bf k})\equiv
\sum_{\alpha \sigma }\left| \phi _{\alpha \sigma }({\bf k)}\right| ^{2}/2$.
The upper and lower signs correspond to $\nu =1$ and 2, respectively. For
the isotropic coupling $\phi ^{2}({\bf k})=1$ a hybridization gap between
the lower and upper hybrid bands $E_{1k}$ and $E_{2k}$ is equal to $\Delta
E\equiv \min E_{2}-\max E_{1}\approx W\left| V_{eff}\right| ^{2}/(W-\mu )\mu 
$ where $W$ is the width of the conduction band. The gap is opened at $%
\varepsilon _{f}$. Calculating the eigenfunctions related to $E_{\nu k}$,
the matrix element $T_{\nu \alpha k,\sigma p}$ can be found:

\begin{equation}
T_{\nu \alpha k,\sigma p}\equiv T_{\alpha k,\sigma p}u_{\nu k}=i\phi
_{\alpha \sigma }({\bf k)}u_{\nu k}T_{k,p}/\phi ({\bf k).}  \label{t-matrix}
\end{equation}
This equation shows that $T_{\nu \alpha k,\sigma p}$ differs strongly from
the bare matrix element $T_{k,p}$.\ The factors $i\phi _{\alpha \sigma }(%
{\bf k)}/\phi ({\bf k)}$ and $u_{\nu k}$ describe the renormalization of the
tunneling matrix element due to the anisotropy of the Kondo coupling and the
formation of the hybrid states (\ref{h-bands}). There is the relation $%
\left| u_{\nu k}\right| ^{2}=\partial E_{\nu k}/\partial \varepsilon _{Rk}$.
The hybrid quasiparticles are called `heavy-fermions' as the mass of the
quasiparticles at the Fermi surface is much larger than the bare electron
mass $m_{R}$: $m_{R}^{\ast }/m_{R}=\left| u_{1k}\right| ^{-2}\gg 1$ at $%
k=k_{F}$. In accordance with Eq. (\ref{t-matrix}) at $T<T_{k}$ the
renormalized matrix element $T_{\nu \alpha k,\sigma p}$ for tunneling into
HF states near the Fermi surface becomes in $1/\left| u_{1k}\right| \sim
(m_{R}^{\ast }/m_{R})^{1/2}$ times smaller than the $T_{k,p}$ value.
Substituting Eq. (\ref{t-matrix}) into Eqs. (\ref{current1}) and (\ref
{power1}) gives

\begin{eqnarray}
I &=&4\pi e\sum_{\nu k,p}\left| T_{k,p}\right| ^{2}\left| u_{\nu k}\right|
^{2}\delta (\varepsilon _{Lp}+eV-E_{\nu k})\times  \nonumber \\
&&[f(E_{\nu k}-eV)-f(E_{\nu k})],  \label{current2} \\
P &=&4\pi \sum_{\nu k,p}\left| T_{k,p}\right| ^{2}\left| u_{\nu k}\right|
^{2}\delta (\varepsilon _{Lp}+eV-E_{\nu k})\times  \nonumber \\
&&(E_{\nu k}-eV-\mu )[f(E_{\nu k}-eV)-f(E_{\nu k})].  \label{power2}
\end{eqnarray}

It is important to note that the increase of the density of states (DOS) in
the hybrid bands, $\rho _{\nu }(E)\equiv \sum_{k}\delta (E-E_{\nu k})=\rho
_{R}\left| u_{\nu k}\right| ^{-2}$, with regard to the bare DOS $\rho _{R}$,
compensates for the decrease of the tunneling matrix element Eq. (\ref
{t-matrix}), as $\rho _{\nu }(E)\left| T_{\nu \alpha k,\sigma p}\right|
^{2}=\rho _{R}\left| T_{k,p}\right| ^{2}$. Taking into account this result,
it is concluded that the formation of heavy fermions does not change
noticeably the magnitude of the tunneling current $I$\ through the NIN$_{HF}$
junction.

In order to examine the properties of the NIN$_{HF}$ junction caused by the
formation of the hybridization gap, the following approximation is made. The
energy dependences of the bare density of states is neglected, as is also
the momentum dependence of $T_{k,p}$ and anisotropy of the Kondo coupling.
With these assumptions the following is obtained

\begin{eqnarray}
I &=&4\pi e\rho _{L}\rho _{R}\left| T_{k,p}\right| ^{2}\sum_{\nu }\int_{\min
E_{\nu }}^{\max E_{\nu }}[f(E-eV)-f(E)]dE,  \nonumber \\
P &=&4\pi \rho _{L}\rho _{R}\left| T_{k,p}\right| ^{2}\sum_{\nu }\int_{\min
E_{\nu }}^{\max E_{\nu }}(E-eV-\mu )\times  \nonumber \\
&&[f(E-eV)-f(E)]dE.  \label{power3}
\end{eqnarray}
At $\left| eV\right| \ll T$ these equations give $P=\Pi I$ where

\begin{eqnarray}
\Pi &=&[\Delta E+E_{t}f(E_{t})-E_{b}f(E_{b})  \nonumber \\
&&-T\ln (f(E_{t})/f(E_{b}))]/(1+f(E_{b})-f(E_{t}))\left| e\right| .
\label{pi}
\end{eqnarray}
Here $E_{t}=\max E_{1k}$, $E_{b}=\max E_{2k}$

In order to obtain the temperature dependence of the Peltier coefficient $%
\Pi $ the set of mean-field equations are solved numerically and $%
\varepsilon _{f}(T)$, $V_{eff}(T)$ and $\mu (T)$ are obtained. These
parameters determine the energy spectrum Eq. (\ref{h-bands}). The
temperature dependences of $\Pi (T)$ and the energy gap $\Delta E$ are
presented in Fig. \ref{figure2}. These calculations show that in the case of a
low-lying Kramers doublet $N=2$ the Peltier coefficient $\Pi (T)$ achieves a
peak value of order of $0.5T_{k}/\left| e\right| $ at $T\propto 0.2T_{k}$.
Therefore, if $eV>0,$ then a current flowing from the left to the right
cools the electrons in the normal metal electrode. The cooling effect has
the following origin. Consider a flow of electrons from the HF electrode
into the normal metal electrode. Such a flow corresponds to $I>0$. In
accordance with the above calculations, electrons with an average energy
smaller than $\mu $ are transferred from the HF electrode into the normal
metal due to the hybridization gap which lies above the chemical potential $%
\mu $ in the energy spectrum of the HF metal, while electrons with an
average energy equal to $\mu $ leave the normal metal electrode through
normal contact.

In the case of impurity scattering at $T<T_{k}$ the Peltier coefficient $\Pi
_{HF}$ of HF metals is equal to $\Pi _{HF}=\pi ^{2}T^{2}N_{f}/(3eN_{c}T_{0})$
where $T_{0}$ is the low temperature Kondo scale, $T_{0}\propto T_{k}$ \cite
{bg91}. In accordance with the calculations represented in Fig. \ref{figure2}
for $N=2$, at $T=T_{m}$ the ratio of the peak value $\Pi (T_{m})$ to $\Pi
_{HF}$ is given by $\Pi (T_{m})/\Pi _{HF}\approx
0.15N_{c}T_{k}^{2}/N_{f}T_{m}^{2}.$ As $N_{c}\sim N_{f}$ and $T_{m}\approx
0.24T_{k}$, $\Pi (T_{m})$ is three times larger than $\Pi _{HF}$. This means
that at low temperatures $T\sim T_{m}$ the NIN$_{HF}$ junction is
substantially more efficient in cooling than the normal metal-heavy fermion
metal contact.

Unlike Ce$^{+3}$ ions, a magnetic moment of Yb$^{+3}$ ions is produced by
one hole in its $f$ shell, and the hybridization gap is opened at $%
\varepsilon _{f}<\mu $. If the right electrode is made from a Yb based HF
compound, then $\Pi $ is negative unlike the positive $\Pi $ for Ce based HF
compounds. In the latter case, electrons in the normal metal thermoelement
are cooled when $eV<0$. For more effective cooling of the normal metal
electrode a junction N$_{HF}$(Yb)ININ$_{HF}$(Ce) may be proposed because the
current flowing through both N$_{HF}$(Yb)IN and NIN$_{HF}$(Ce) junctions
will cool the electrons in the normal metal electrode.

The above calculations of the Peltier coefficient of NIN$_{HF}$ junctions
have been carried out in the isotropic approximation neglecting anisotropy
of the Kondo coupling produced by the crystal field effect. It is well known
that this effect influences the DOS in HF compounds \cite{Ikeda96,mc00}. It
is also expected that it will affect the tunneling current and the power
transfer through the junction.

In conclusion, the tunneling current and Peltier effect of the normal
metal-insulator-heavy fermion metal junctions have been studied. It has been
demonstrated that at temperatures below the Kondo temperature the tunneling
current transfers energy from the normal metal cooling electrons in the
normal metal electrode. Thus, these junctions as well as the
normal-insulator-superconductor junctions may be used as an electronic
microrefrigerator to cool electrons in the normal metal electrode below the
lattice temperature.

This work was supported by the EPSRC (UK) grant No. 04003/01 and in part by
the Russian Fund for Basic Research, Grant No.01-02-17794.

\begin{figure}[tbp]
\caption{(a) Schematic of the NIN$_{HF}$ junction. (b) The electron
distribution function in the normal metal electrode (on the left) and HF
electrode (on the right).}
\label{figure1}
\end{figure}

\begin{figure}[tbp]
\caption{(a) The hybridization gap $\Delta E$ versus $T$ in the HF metal for
the spin degeneracy $N=2$ and concentration of conduction electrons $%
N_{c}=0.95$ per $f$ ion. (b) Temperature behavior of the Peltier coefficient 
$\Pi $ for the NIN$_{HF}$ junction determined from Eq. (\ref{pi}) at $N=2$
(solid line), $N=8$ (dashed line). }
\label{figure2}
\end{figure}

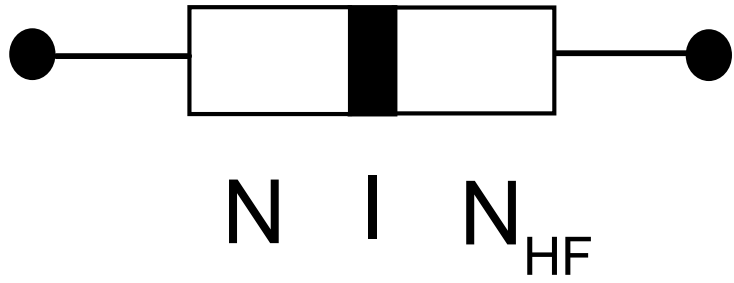
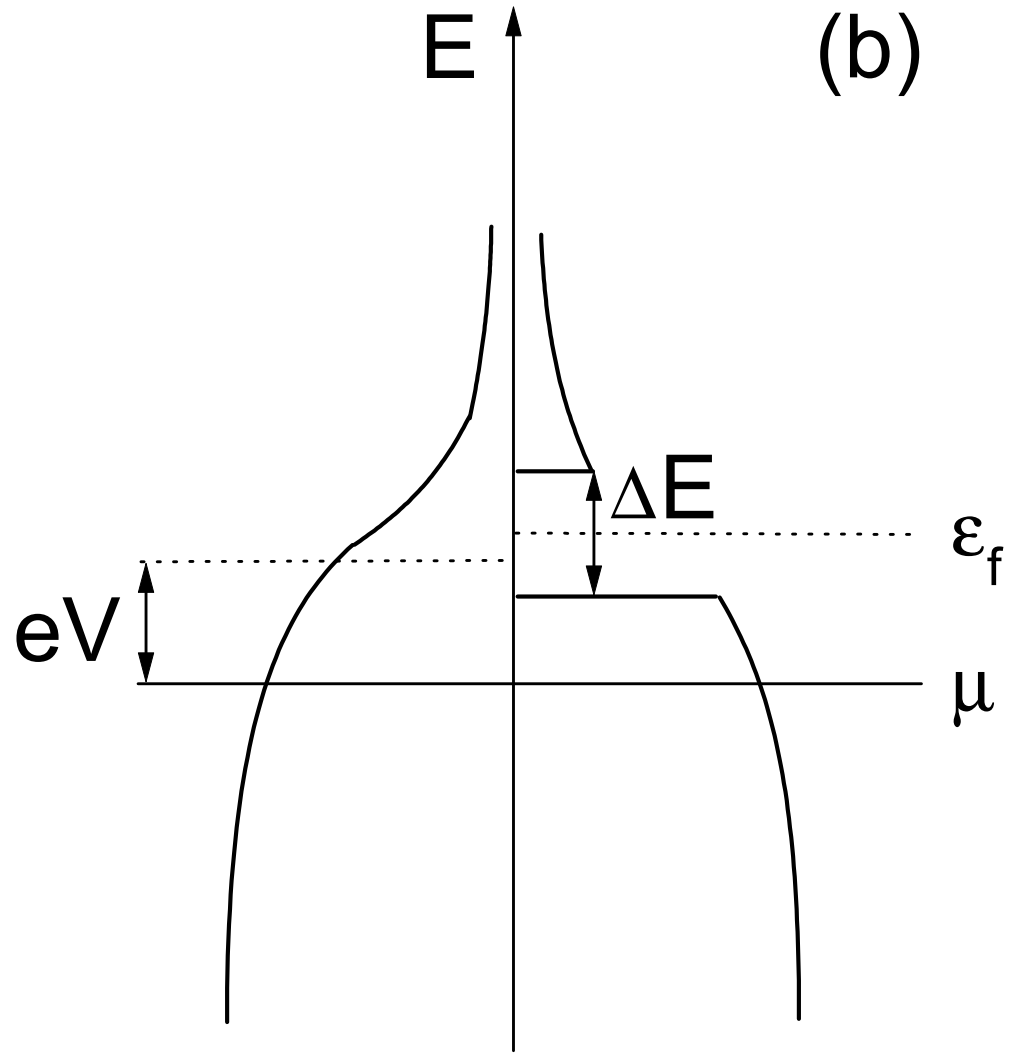
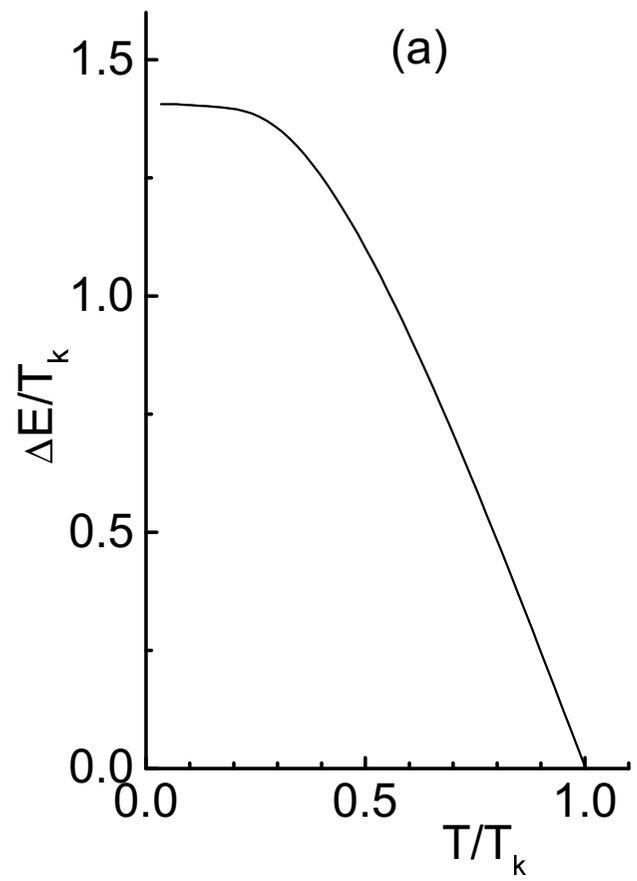
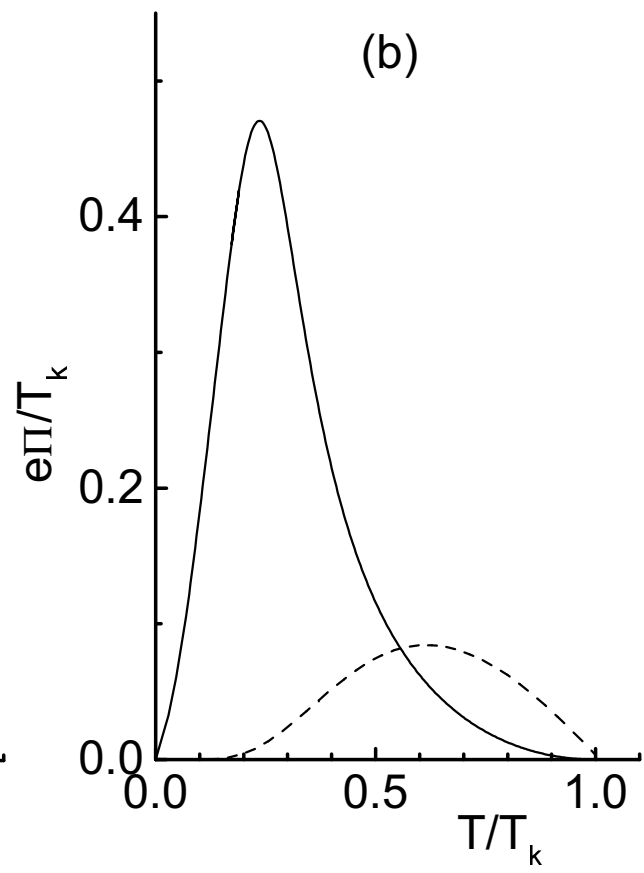
\end{document}